\documentclass[oldversion]{aa} 
\usepackage{amsmath}
\usepackage{natbib,txfonts} 
\usepackage{graphicx} 
\bibpunct{(}{)}{;}{a}{}{,}

\def\ho{\mathrm{H}_{0}}

\def\lx{\mathrm{L}_{\mathrm{X}}}
\def\kt{\mathrm{kT}} 
\def\ab{\mathrm{Z}} 
\def\kto{\mathrm{kT_o}} 
\def\abo{\mathrm{Z_o}} 
\def\T{\mathrm{T}}
\def\kev{\mathrm{keV}}

\def\w{\mathrm{W}}

\def\msol{\mathrm{M}_{\odot}}
 
\def\nh{\mathrm{N}_\mathrm{H}}
\def\nho{\mathrm{N}_\mathrm{H,o}}

\def\sx{\mathrm{\Sigma}_\mathrm{x}}  
\def\yx{\mathrm{Y}_{\mathrm{X}}}

\def\a2163{A2163}
\def\xmm{XMM-Newton~}

\def\f{\mathrm{F_{\mathrm{evt}}}}
\def\s{\mathrm{F_{\mathrm{ICM}}}} 
\def\c{\mathrm{F_{\mathrm{gal,CXB}}}}
\def\b{\mathrm{F_{\mathrm{bck}}}} 
\def\o{\mathrm{F_{\mathrm{oot}}}} 
\def\p{\mathrm{F_{\mathrm{p}}}}
\def\nf{\mathrm{n}_\mathrm{evt}} 
\def\ns{\mathrm{n}_\mathrm{ICM}}
\def\nc{\mathrm{n}_\mathrm{gal,CXB}} 
\def\nb{\mathrm{n}_\mathrm{bck}}
\def\np{\mathrm{n}_\mathrm{p}} 
\def\no{\mathrm{n}_\mathrm{oot}}
\def\ea{\mathrm{E}}

\def\Theta{\mathrm{Theta}}

\def\fig{Fig.~}
\def\tab{Table~}
\def\equ{Eq.~}

\def\part{Sect.~}

\begin{document}

\title{A2163: Merger events in the hottest Abell galaxy cluster}
\subtitle{II. Subcluster accretion with galaxy-gas separation}
\titlerunning{A2163: Merger events in the hottest Abell galaxy cluster. II. ICM dynamics}

\author{H. Bourdin \inst{1}, M. Arnaud\inst{2}, P. Mazzotta\inst{1}, G.W. Pratt\inst{2}, J.-L. Sauvageot\inst{2}, R. Martino \inst{1}, S. Maurogordato\inst{3}, A. Cappi\inst{3,4}, C.~Ferrari\inst{3}, and C. Benoist\inst{3}}
\authorrunning{Bourdin et al.}
\offprints{H. Bourdin}

\institute{$^1$ Dipartimento di Fisica, Universit\`a degli Studi di Roma `Tor
  Vergata', via della Ricerca Scientifica, 1, I-00133 Roma, Italy \\
$^2$ Laboratoire AIM, IRFU/Service d'Astrophysique - CEA/DSM - CNRS - Universit\'{e} Paris Diderot, Orme des Merisiers B\^{a}t. 709, CEA-Saclay, F-91191 Gif-sur-Yvette Cedex, France \\
  $^3$ Universit\'e de Nice Sophia-Antipolis, CNRS,
  Observatoire de la C\^ote d'Azur,  UMR 6202 CASSIOPEE, BP 4229, F-06304 Nice Cedex 4, France \\
  $^4$ INAF - Osservatorio Astronomico di Bologna, via Ranzani 1, I-40127
  Bologna, Italy \\}

\date{Received 3 May 2010 / Accepted 22 September 2010}

\abstract{Located at $z=0.203$, \a2163 is a rich galaxy cluster with an intra-cluster medium (ICM) that exhibits extraordinary properties, including an exceptionally high X-ray luminosity, average temperature, and a powerful and extended radio halo. The irregular and complex morphology of its gas and galaxy structure suggests that this cluster has recently undergone major merger events that involve two or more cluster components. In this paper, we study the gas structure and dynamics by means of spectral-imaging analysis of X-ray data obtained from XMM-Newton and Chandra observations. From the evidence of a cold front, we infer the westward motion of a cool core across the E-W elongated atmosphere of the main cluster A2163-A. Located close to a galaxy over-density, this gas `bullet' appears to have been spatially separated from its galaxy (and presumably dark matter component) as a result of high-velocity accretion. From gas brightness and temperature profile analysis performed in two opposite regions of the main cluster, we show that the ICM has been adiabatically compressed behind the crossing `bullet' possibly because of shock heating, leading to a strong departure of the ICM from hydrostatic equilibrium in this region. Assuming that the mass estimated from the $\yx$ proxy best indicates the overall mass of the system and that the western cluster sector is in approximate hydrostatic equilibrium before subcluster accretion, we  infer a merger scenario between two subunits of mass ratio 1:4, leading to a present total system mass of $M_{500} \simeq 1.9 \times10^{15}~\msol$. Additional analysis of the spatially-separated northern subcluster A2163-B does not show any evidence of strong interaction  with the main cluster A2163-A, leading us to infer that the physical distance separating the northern subcluster and the main component is longer than the projected separation of these components. The exceptional properties of \a2163 present various similarities with those of 1E0657-56, the so-called `bullet-cluster'. These similarities are likely to be related to a comparable merger scenario.}



\maketitle

\section{Introduction}

\a2163 is a rich galaxy cluster (richness class 2) located at $z=0.203$. After initial X-ray detection by HEAO 1 A-1 \citep{Kowalski_84}, combined observations by the Ginga and Einstein X-ray satellites revealed the extraordinary properties of its hot gas content, with exceptionally high luminosity and average temperature \citep[$\lx = 3.5 \times 10^{38}\, \w$, $\kt = 13.9\, \kev$,][]{Arnaud_92}\footnote{$10^{38}\, \w \equiv 10^{45} \mathrm{erg}~\mathrm{s}^{-1}$; X-ray luminosity in the 2-10 keV band has been corrected for luminosity distance assuming $\ho = 70$ km s$^{-1}$ Mpc$^{-1}$, and $\Lambda = 0.7$.}, suggesting a very high cluster mass. Extensive follow-up observations have further revealed several signatures of major merger events at various wavelengths in this hottest Abell cluster. ROSAT observations showed an irregular X-ray morphology \citep{Elbaz_95}, while ASCA observations revealed complex thermal structure \citep{Markevitch_94}, more recently confirmed from Chandra data \citep{Markevitch_01, Govoni_04, Owers_09}. Evidence of a clear cluster substructure has been seen in weak lensing \citep[see e.g.][]{Squires_97,Radovich_08} and in the galaxy density distribution \citep{Maurogordato_08}. Moreover, \a2163 is known to host a prominent radio halo \citep{Feretti_01,Feretti_04} and to be a possible source of non thermal hard X-ray emission, as suggested by \citet{Rephaeli_06} from RXTE observations and \citet{Million_09} from Chandra data analysis. Due to its high thermal energy content, \a2163 is also a favoured target for SZ observations. First detected by \citet{Wilbanks_94} and used to determine the Hubble constant in \citet{Holzapfel_97}, \a2163 is the first galaxy cluster detected in the submm band \citep{Lamarre_98}. A recent analysis of its SZ signal by \citet{Nord_09} has confirmed the exceptionally high temperature of its hot gas content independently of X-ray analyses. 


\begin{figure*}[ht]
    \resizebox{.95\hsize}{!}{\includegraphics{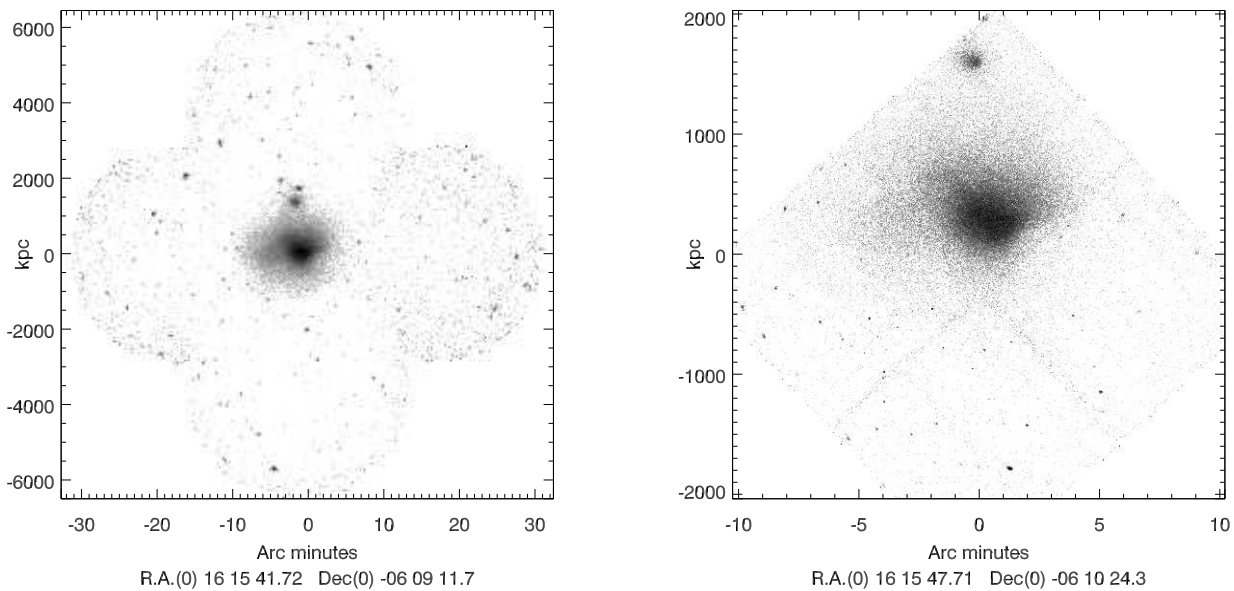}}
  \caption{\xmm-EPIC exposure mosaic (left) and Chandra-ACIS exposure (right).\label{exposure_fig}}
\end{figure*}

Quite apart from the evidence for a very high total system mass, the exceptionally large amount of thermal energy trapped within the hot atmosphere of \a2163 is likely to be related to ongoing cluster merger events. For this reason, the mass of \a2163 has been challenging to estimate from X-ray analyses, which rely on the gas being in hydrostatic equilibrium. Indeed, departure of the cluster gas content from hydrostatic equilibrium may explain the 50 \% dispersion observed between mass estimates of  e.g. \citet{Elbaz_95} and \citet{Markevitch_96}, performed assuming different temperature gradients in the cluster outskirts. It is also interesting to note the discrepancy between X-ray measurements of the gas temperature and the indirect weak lensing temperature estimate of \citet{Cypriano_04}, $kT_{SIS} = 6.63^{+2.03}_{-1.76}$ keV, obtained assuming energy equipartition between galaxies and gas.

Our recent analysis of optical data (WFI/2.2m, VIMOS/VLT, UT3) in \citet[][hereafter Paper I]{Maurogordato_08} revealed the complex galaxy structure and dynamics in \a2163, where a main cluster, \a2163-A, is connected to a smaller subcluster to the North, \a2163-B, via a galaxy bridge. The main cluster, A2163-A, itself has two brightest galaxies (BCG1 and BCG2), a bimodal morphology, and an exceptionally large velocity dispersion ($1434\pm60$ km s$^{-1}$). As part of a multi-wavelength analysis programme aimed at investigating the impact of cluster mergers on galaxy and ICM evolution, the present article focuses on the analysis of \xmm and Chandra observations of the ICM in \a2163. This work particularly aims to investigate the merger scenario involving the various components of \a2163, and to provide new insights on the merger impact on X-ray mass estimates. Data preparation and analysis issues related to the \xmm and Chandra observations are detailed in Sections \ref{obs_dataprep} and \ref{data_analysis}. Results obtained on ICM substructure and merger dynamics are discussed in Section \ref{icm_dynamics}, while Sections \ref{icm_pressure_sct} and \ref{mass_sct} concern the gas pressure structure and mass of \a2163. Discussion and concluding remarks are provided in Section \ref{conclusion}. Except if otherwise noted, the confidence ranges on individual parameter estimates are 68 $\%$. In the following, $\mathrm{M}_\delta$ is defined as the mass within the radius $\mathrm{r}_\delta$ at which the mean mass density is $\delta$ times the critical density of the universe at the cluster redshift. Moreover, intra-cluster distances are computed as angular diameter distances, assuming a $\Lambda$-CDM cosmology with $\mathrm{H}_\mathrm{0} =70~\mathrm{km}~\mathrm{s}^{-1}~\mathrm{Mpc}^{-1}$, $\Omega_{\mathrm{M}} = 0.3$, $\Omega_{\Lambda} = 0.7$. Given these assumptions, an angular separation of 30 arcsec corresponds to a projected intra-cluster distance of 100 kpc.

\begin{table*}[ht]
\caption{Effective exposure time of each \xmm{}-EPIC and Chandra-ACIS observation.  In
brackets: fraction of the useful exposure time after solar-flare
``cleaning''. \label{pointing_params_tab}}
\begin{center}
\begin{tabular}{ccccc}
\hline\hline
XMM-Newton  & Centre coordinates & MOS1  effective  &  MOS2  effective &  PN  effective \\
obs. IDs & & exposure time (ks) & exposure time (ks) & exposure time (ks)  \\
\hline
0112230601  &  16h15m46.01s -06$^\circ$09'00.0" & 8.8 (56.2  \%)  & 9.8  (63.5 \%) &  5.7  (56.2 \%) \\
0112230701  & 16h16m48.00s -06$^\circ$09'00.0" &11.8 (42.5  \%)  & 12.1  (43.4 \%) &  6.1  (29.9 \%) \\
0112230801  & 16h15m46.01s -05$^\circ$51'00.0" &16.1 (57.5  \%)  & 14.6   (52.2\%) & 10.2  (49.0 \%) \\
0112230901  &  16h14m34.01s -06$^\circ$09'00.0" &7.4 (27.5  \%)  &  6.9  (25.5 \%) &  3.3  (16.7 \%) \\
01122301001 & 16h15m46.01s -06$^\circ$27'00.0" &19.5 (69.3 \%)  & 16.5   (58.8 \%) & 15.4  (72.4 \%) \\
\hline
Chandra obs ID & Centre coordinates & \multicolumn{3}{c}{ACIS effective exposure time (ks)} \\
\hline
01653 & 16h15m45.77s -06$^\circ$ 08'55.0" & \multicolumn{3}{c}{65.2 (91.7 \%)} \\
\hline
\end{tabular}
\end{center}
\end{table*}

\section{Observations and data preparation\label{obs_dataprep}}

\subsection{Observations}

\subsubsection{The XMM-Newton data set}
 
The \xmm{} data set is a mosaic composed of five observations obtained with the European Photon Imaging Camera (EPIC), and available in the \xmm{} archive (observation IDs: 0112230601, 0112230701, 0112230801, 0112230901, 0112231001). This data set has been used to map the ICM thermal structure in the central region of \a2163, and to investigate its radial brightness and temperature structure on a large scale. Following a similar scheme as in \cite{Bourdin_08}, hereafter BM08, we filtered the data set through spatial and temporal wavelet analyses in order to remove the contribution of point sources and soft proton flares. A summary of the effective exposure time remaining after filtering is provided in \tab \ref{pointing_params_tab}.

\subsubsection{The Chandra data set}

The Chandra data set is a 72 ks pointing observation of \a2163 obtained with the AXAF-I CCD Imaging Spectrometer (ACIS-I) and available in the Chandra archive (observation ID: 01653). These data have been used to map the gas brightness structure at high angular resolution near the cluster centre. Following a similar procedure as for the XMM-Newton data, we filter events in order to remove point source and soft proton flare contributions (see effective exposure in Tab. \ref{pointing_params_tab}).

\subsection{Data preparation}

In order to perform self-consistent brightness measurements and spectroscopic estimates using the \xmm or Chandra data sets (see Section \ref{data_analysis}), we re-sampled photon events within fixed grids matching the angular and spectral resolution of each focal instrument in sky coordinates $(k,l)$ and energy $(e)$. Events associated with the five \xmm pointings have thus been gathered into three event cubes corresponding to each of the three EPIC instruments, while events associated with the Chandra pointing have been binned into a unique event cube. 
Following a similar procedure as described in BM08, we then associated a set of local `effective exposure' and `background noise' arrays  to these event cubes. The effective exposure $\ea(k,l,e)$ is computed as a linear combination of exposure times associated with each pointing, with correction for spatially variable mirror effective areas, filter and other focal instrument transmissions, CCD pixel area with correction for telescope motion, gaps and bad pixels\footnote{Information about these instrumental effects have been obtained from the \xmm-EPIC Current Calibration Files (CCFs) and Chandra-ACIS Calibration database (CALDB 4.1)}. The `background noise'  array $\b(k,l,e)$ has been modelled as the sum of 3D functions accounting for galaxy foreground and Cosmic X-ray Background (CXB) emissivities, $\c(e)$, and also false photon detection from the particle induced and out-of-time events, $\p(k,l,e)$ and $\o(k,l,e)$, respectively\footnote{In \equ(\ref{background_equ}) and (\ref{brightness_estimate}), spectral models are decomposed into normalised 
energy distributions, $\mathrm{F}_i(e)$, and counts per pixel $\mathrm{n}_i$\label{equ_12_footnote}}:

\begin{eqnarray}
  \nb(k,l) \b(k,l,e) &=& ~ \ea(k,l,e) \times \nc~\c(e) \nonumber \\ 
  &&+ \no(k,l)~\o(k,l,e) \nonumber \\
  &&+ \np(k,l)~\p(e).
  \label{background_equ}
\end{eqnarray}

\begin{figure}[t]
    \resizebox{\hsize}{!}{\includegraphics{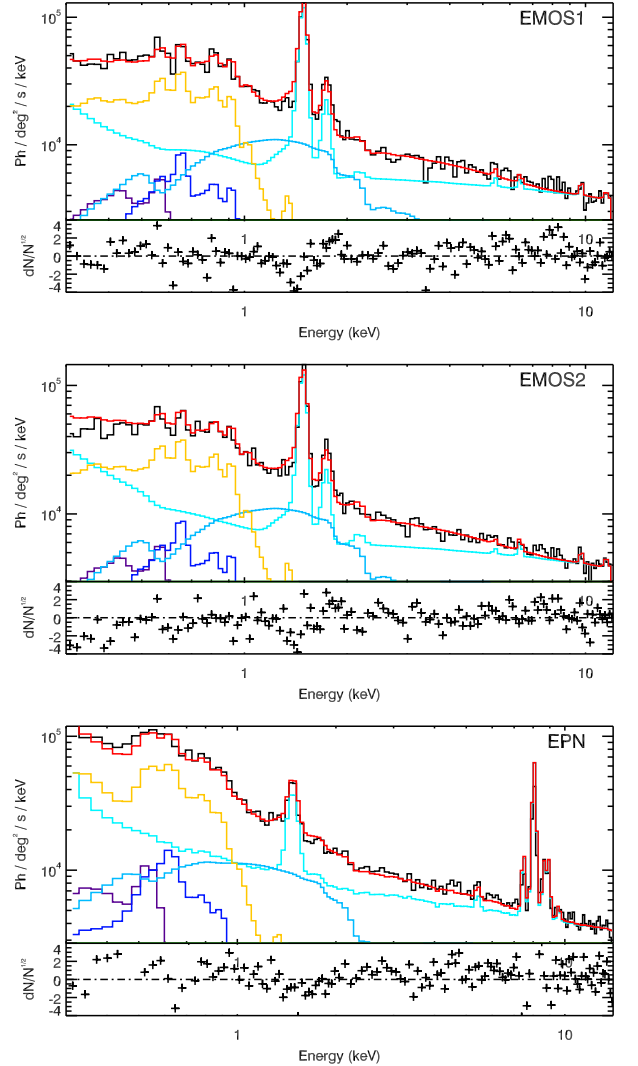}}
  \caption{Background spectrum observable in the A2163 outskirts (r$>$3Mpc) from the four external
    pointings in \tab\ref{pointing_params_tab}. Light blue: particle background noise. Cyan blue: 
    Cosmic X-ray Background emission. Blue and Violet: TAE emission \citep[$\mathrm{kT}_1=0.099~\kev$, $	
    \mathrm{kT}_2= 0.248~\kev$, see][and details in \part\ref{foreground_background_sct}]{Kuntz_00}. 
    Orange: Local Hot Bubble and North Polar Spur residue. Red and black: overall fit and data set.\label{background_fig}}
\end{figure}

As described in \citet{Bourdin_04}, 
we model the particle induced background $\p(e)$ as the sum of a dual power-law continuum and a set of detector fluorescence lines (see light blue curves on \fig\ref{background_fig}). This model has been fitted to a set of observations obtained during `closed' periods of the \xmm telescopes, and stowed periods of the Chandra ACIS detector. For both the \xmm and Chandra observations, the galaxy foreground and CXB emissivities, $\c(e)$, have been estimated within an external annulus located away from the target emission using offset pointings of the EPIC-\xmm mosaic. Further details concerning our modelling of the extended foreground and background emissions are provided in Section \ref{foreground_background_sct}.

\section{Data analysis\label{data_analysis}}

\subsection{X-ray brightness measurements and spectroscopy}

The event, `effective exposure', $\ea(k,l,e)$, and `background noise', $\b(k,l,e)$, arrays have enabled us to perform
ICM brightness and average temperature measurements in various areas of the field of view by gathering pixels $(k,l)$. 

We computed brightness estimates $\sx(k,l)$ from photon counts $\nf(k,l,e)$ in the `soft' energy band (0.5-2.5 keV) 
as follows$^{\ref{equ_12_footnote}}$:

\begin{equation}
		\sx(k,l) = \frac{\sum_{e}{\nf(k,l,e)-\nb(k,l) \sum_{e}\b(k,l,e)}}{\sum_{e}\s(\kto,\abo,\nho,e)\ea(k,l,e)},
	\label{brightness_estimate}
\end{equation}

where $\s(\kto,\abo,\nho,e)$ is an emission model associated with average source temperature, 
$\kto$, metallicity, $\abo$, and Galactic hydrogen absorption $\nho$. Hereafter, the source emission spectrum $\s(\kt,\ab,\nh,e)$ assumes an $\nh$ absorbed emission modelled from the Astrophysical Plasma Emission Code \citep[APEC,][]{Smith_01}, with the element abundances of \citet{Grevesse_98} and neutral hydrogen absorption cross sections of \citet{Balucinska-Church_92}. It is altered by a local energy response matrix M(k,l,e,e') computed from response matrixes files (RMF) tabulated in detector coordinates in the \xmm-EPIC and Chandra-ACIS calibration data bases \footnote{EPIC response matrixes are computed from canned RMFs corresponding to the observation period provided by the \xmm Science Operation Centre. ACIS responses matrixes have been computed using the Chandra Interactive Analysis of Observations (CIAO) software and Chandra-ACIS Calibration database (CALDB 4.1)}. 

Following this emission model, average temperatures $\kt(k,l)$ are computed from spectral fitting to the emission spectrum $\nf(k,l)\f(k,l,e)$, registered in the energy band (0.7-12 keV):

\begin{eqnarray}
  && \nf(k,l)~\f(\kt,\ab,\nh,e)  \nonumber \\
  &=& \ea(k,l,e) \times \ns(k,l)~\s(\kt,\ab,\nh,e) \nonumber \\
  && + \nb~\b(k,l,e).
\label{temperature_estimate}
\end{eqnarray}

\begin{figure*}[t]
    \resizebox{.95\hsize}{!}{\includegraphics{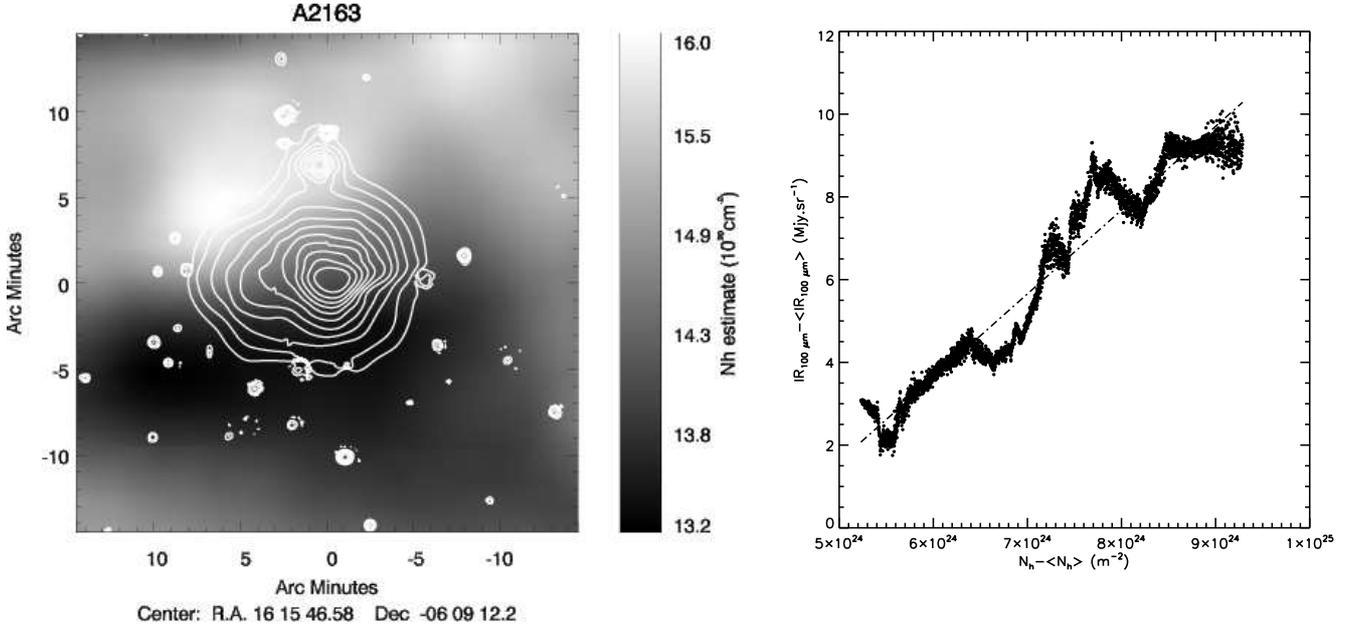}}
  \caption{Left: Galactic neutral hydrogen density column estimated in the neighbourhood of \a2163, with contours of the [0.5-2] keV \xmm-EPIC emission overlaid. Right: Correlation factor, $\frac{\nh}{\tau_{\mathrm{IR}}}$, relating IR 
	emissivity to the Galactic $\nh$ in a 5 square degree region of the sky centred on \a2163. IR emissivity and Galactic $\nh$
	have been corrected for dependence on galactic latitude following the analytic approach of \citet{Desert_88}. A correlation factor $\frac{\nh} {\tau_{\mathrm{IR}}}$ is computed between the average residual IR emissivity and Galactic $\nh$ within sky regions corresponding to a fixed $\nh$ interval of width $\delta \nh = 10^{21} m^{-2}$ \label{nh_map_fig}.}
\end{figure*}

\subsection{X-ray brightness, ICM density and temperature profiles}

Through the above analysis, X-ray brightness and temperature profiles have been extracted assuming uniform emissivity
within concentric annuli. We derived the radial brightness $\sx(r)$ and associated Gaussian fluctuation $\sigma_{\sx}(r)$
by averaging the local brightnesses $\sx(k,l)$ of \equ(\ref{brightness_estimate}) in each profile annulus composed of N 
pixels $(k,l)$, as follows:

\begin{eqnarray}
	\sx(r) &=& \frac{1}{N}\sum_{k,l} \sx(k,l) \\
\sigma_{\sx}(r) &=& \frac{1}{N}\sqrt {\sum_{k,l}\frac{\sum_{e}\nf(k,l,e)}{\left[\sum_{e}\s(\kto,\abo,\nho,e)\ea(k,l,e)\right]^2}}.
\end{eqnarray}

Radial temperatures $\kt(r)$ and associated confidence interval $\delta \kt(r)$ have been computed within each annulus by fitting a uniform emission model to the data set. To do so, we averaged emission models of \equ(\ref{temperature_estimate}) associated with each pixel $(k,l)$ of the annulus, and fitted an average set of spectroscopic parameters $\kt(r),\ab(r),\nh(r)$ by means of a $\chi^2$ minimisation. 

These brightness and temperature profiles have been used to model the underlying density and temperature of the ICM, 
assuming spherical symmetry of the cluster atmosphere. This was undertaken by projecting and fitting parametric distributions of the radial emission measure, $n_p n_e$, and temperature, $\T(r)$, to the observed profiles. Introduced by \citet{Vikhlinin_06}, these parametric distributions are:

\begin{eqnarray}
  [n_p n_e](r) &=& n_{0}^2
  \frac{(r/r_c)^{-\alpha}}{[1+(r/r_c)^2]^{3\beta - \alpha/2}}
  \frac{1}{[1+(r/r_s)^3]^{\epsilon/3}} \nonumber \\ &&+ \frac
  {n_{02}^2}{[1+(r/r_{c2})^2]^{3\beta_2}},
  \label{rho3d_equ}
\end{eqnarray}

\begin{equation}
  \T(r) = \T_o \frac{(r/r_t)^{-a}}{[1+(r/r_t)^b]^{c/b}},
  \label{t3d_equ}
\end{equation}

In \part\ref{moving_cc}, the ICM emission measure and temperature profiles have been modelled by two step distributions with common jump radius $r_j$, in order to calculate the physical characteristics of a cold front in the cluster atmosphere:

\begin{equation}
  [n_p n_e]_{cf}(r) = \left\{ \begin{matrix}
    D_{n}^2 n_o^2 (r/r_{j})^{-2\eta_1}
    \left[ \frac {1+\left(r_{j}/r_{c1}\right)^2} {1+\left(r/r_{c1}\right)^2} \right]^{3\beta_1}, 
    ~r<r_{j} \\
    n_o^2 (r/r_{j})^{-2\eta_2}
    \left[ \frac {1+\left(r_{j}/r_{c2}\right)^2} {1+\left(r/r_{c2}\right)^2} \right]^{3\beta_2}, 
    ~r>r_{j}
  \end{matrix} \right. ,
  \label{npne_cf_equ}
\end{equation}

\begin{equation}
  \T_{cf}(r) = \left\{ \begin{matrix}
    \T_o, ~~r<r_{j} \\
	D_{\T} \T_o, r>r_{j} \end{matrix} \right. .
  \label{t3d_cf_equ}
\end{equation}

\subsection{Imaging and spectral-imaging\label{lxkt_maps}}

\subsubsection{Imaging}

The ACIS-Chandra data set has enabled us to analyse the ICM brightness structure in \a2163 at the highest angular resolution available. By means of two types of wavelet analysis, we first mapped the ICM brightness of the large-scale cluster emission, then isolated the brightness structure of the central region from contributions related to the cluster outskirts.

To map the large-scale ICM brightness, we first corrected `soft' photon (0.5-2.5 keV) counts for vignetting and background emissivity as detailed in \equ(\ref{brightness_estimate}), then analysed the corrected image following the redundant `a-trous' algorithm \citep{Holschneider_89, Shensa_92} implemented with B3-spline wavelets. A de-noised brightness map was then constructed from the overall wavelet transform of the image, with coefficients thresholded iteratively assuming that photon counts follow a Poisson distribution \citep[see][and references therein for details]{Bijaoui_95,Starck_98b}. In the top-left panel of \fig\ref{lxkt_maps_fig}, this map is overlaid on an $R$-band optical image obtained at the WFI/MPI 2.2m telescope (see also Paper I).

To further isolate the ICM brightness structure in the cluster centre, we restricted a B3-spline wavelet analysis of a corrected photon image extracted in the [0.5-7.5] keV energy band to four high resolution scales, corresponding to a distance range of [1 - 120 arcsec]. The resulting high resolution residue of the ICM brightness map shown on \fig\ref{lxkt_maps_fig} has been constructed by summing details of this restricted wavelet transform, with coefficients thresholded to a 4 $\sigma$ confidence level.

\subsubsection{Spectral-imaging}

\begin{figure*}[ht]
  \resizebox{.98\hsize}{!}{\includegraphics{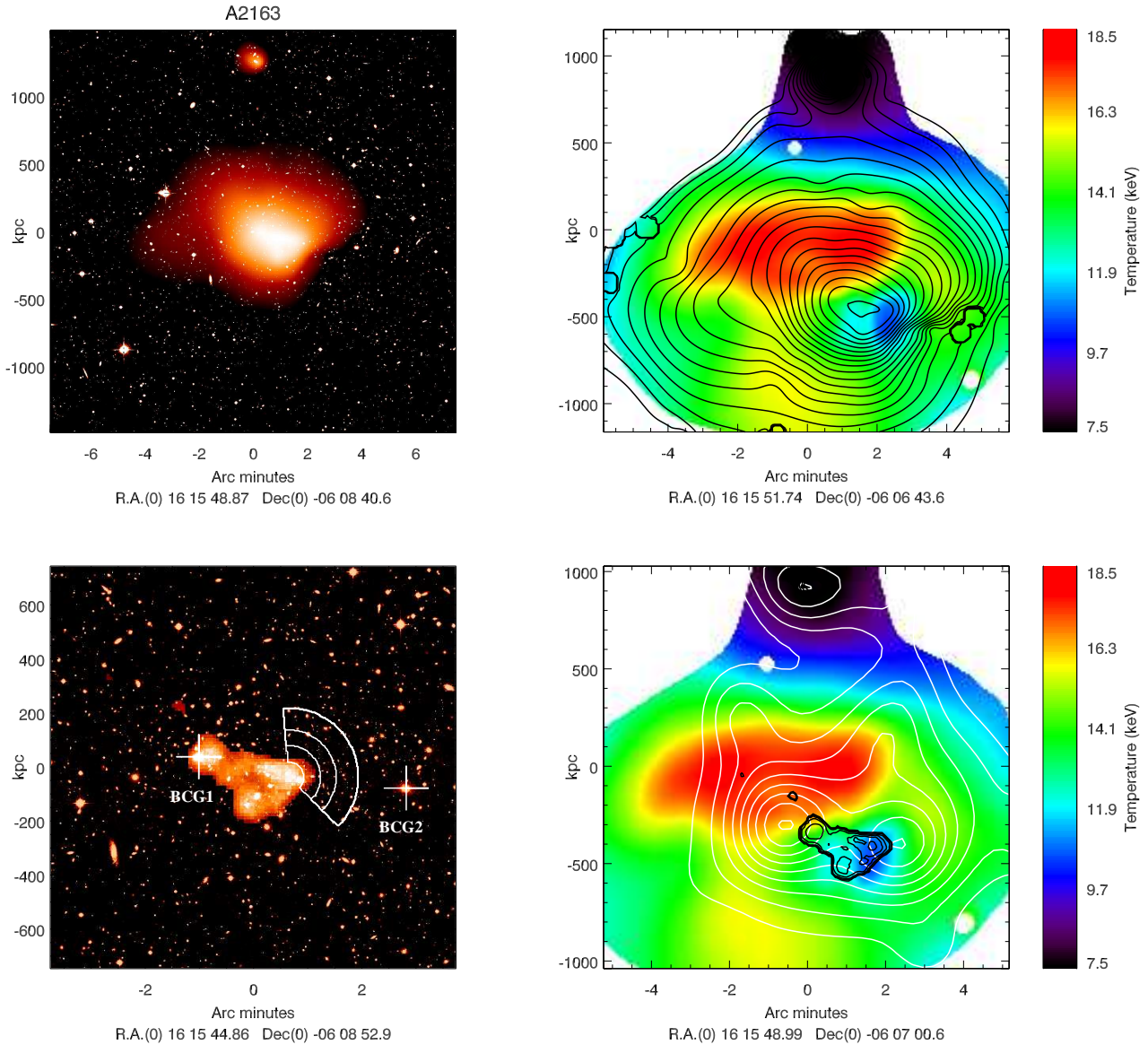}}
  \caption{A2163. Left panels: X-ray emission overlaid on galaxy maps. Optical observations performed at the MPI 2.2m telescope, see Paper I \citep{Maurogordato_08}. X-ray images obtained from wavelet analyses of a Chandra exposure in the 0.5-2.0 keV band. Top-left: overall cluster emission. Bottom-left: high resolution 
	analysis of the ICM emission near the cluster centre, see Section \ref{lxkt_maps} for details. Right panels: ICM temperature maps
	obtained from wavelet spectral-imaging analysis of the EPIC-\xmm data set. Top-right: black iso-contours from the Chandra image (same as top-left map) overlaid as black iso-contours. Bottom-right: black iso-contours from high resolution Chandra residue (same as bottom-left map); white iso-density contours from galaxy density map (see also \fig 6 in Paper I). \label{lxkt_maps_fig}}
\end{figure*}

In order to map the ICM temperature in \a2163, we used the EPIC-\xmm data set and applied the spectral-imaging algorithm detailed in \citet{Bourdin_04} and B08. Following this algorithm, a set of temperature arrays $\kt(k,l,a)$ with associated fluctuations  $\sigma_{kt}(k,l,a)$ are first computed on various analysis scales $a$, then convolved by complementary high-pass and low-pass analysis filters in order to derive wavelet coefficients. The wavelet coefficients are subsequently thresholded according to a given confidence level in order to restore a de-noised temperature map. Here, the signal analysis have been performed over 6 dyadic scales within an angular resolution range of $\delta a$ = [3.4 - 220] arcsec. This was undertaken by averaging the emission modelled by \equ(\ref{temperature_estimate}) within overlapping meta-pixels $(k,l,a)$, and computing the $\kt(k,l,a)$ and $\sigma_{kt}(k,l,a)$ arrays by means of a likelihood maximisation assuming the spatially variable Galactic Nh and foreground emission model discussed below in Sect.~\ref{Nh_section}. The resulting ICM temperature map shown in \fig\ref{lxkt_maps_fig} was then obtained from a B2-spline wavelet analysis (see B08 for details) with coefficients thresholded to the 1 $\sigma$ confidence level.
    
\subsubsection{Absorption by Galactic neutral hydrogen \label{Nh_section}}

Located close to the galactic plane, in the direction of the galactic centre (l=19.1$^{\circ}$, b=37.2$^{\circ}$),
\a2163 is placed behind a relatively strong column density of galactic neutral hydrogen, $\nh \simeq 1.5 \times 10^{21}$ cm$^{-2}$. In this 
high $\nh$ regime, temperature measurements are highly sensitive to spatial variations of the Galactic $\nh$ on angular scales
smaller than the cluster size, so that a precise modelling of these variations is required to map the ICM temperature. Indeed, as formerly shown by a ROSAT data analysis of \citet{Elbaz_95}, the combination of the high average temperature of the ICM in \a2163 with this high (and variable) $\nh$ value makes temperature measurements highly degenerate with $\nh$ estimates. For these reasons, we decided to model a priori the spatial variations of neutral hydrogen Galactic absorption across the cluster field of view. 

To investigate the spatial variations of Galactic Hydrogen density column up to an angular resolution of about one arcminute, we used Galactic dust emissivity maps at 100 $\mu$m. The spatial variations of the overall Hydrogen (i.e., both neutral HI, and molecular H$_2$) density column are indeed spatially correlated with galactic dust emissivity, $\tau_{\mathrm{IR}}$, in the 100 $\mu$m IR band \citep[see e.g.][]{Boulanger_96,Snowden_98}. Moreover, as \a2163 is located outside any Galactic molecular cloud \footnote{See e.g. the large-scale CO survey of the Galactic plane by \citet{Dame_01}.}, the global budget of Hydrogen density column is dominated by neutral hydrogen in this region of the sky, enabling us to use galactic dust emissivity maps at 100 $\mu$m as a tracer of $\nh$ spatial variations. To model these variations, we follow a two step approach where large scale variations related to changes in the Galactic Interstellar Radiation Field are separated from smaller scale fluctuations. We used the  IRAS/IRIS \citep{Miville_Deschenes_05} and Leiden/Argentine/Bonn \citep[LAB, ][]{Kalberla_05} surveys of the all sky IR emissivity and Galactic $\nh$, and corrected their dependence with galactic latitude, b, by subtracting two cosecant laws with shape $\frac{1}{sin|\mathrm{b}|}$, following the analytic approach of \citet{Boulanger_88} and equations (3.1) and (3.2) of \citet{Desert_88}. We then measured the large-scale correlation factor, $\frac{\nh}{\tau_{\mathrm{IR}}}$, relating the residual IR emissivity and Galactic $\nh$ in a 5 square degree region of the sky centred on \a2163 (see also \fig\ref{nh_map_fig}). We finally model the smaller scale spatial variations of $\nh$ across the cluster field of view, by applying the $\frac{\nh}{\tau_{\mathrm{IR}}}$ factor to the IRAS/IRIS maps, and deducing the expected $\nh$ map at high angular resolution. As shown in Fig. \ref{nh_map_fig}, this map reveals us $\nh$ variations of the order of 20 \%, with a NorthEast-SouthWest gradient that could have strongly distorted X-ray spectroscopic estimates, if not taken into account.

\begin{figure*}[ht]
  \resizebox{.95\hsize}{!}{\includegraphics{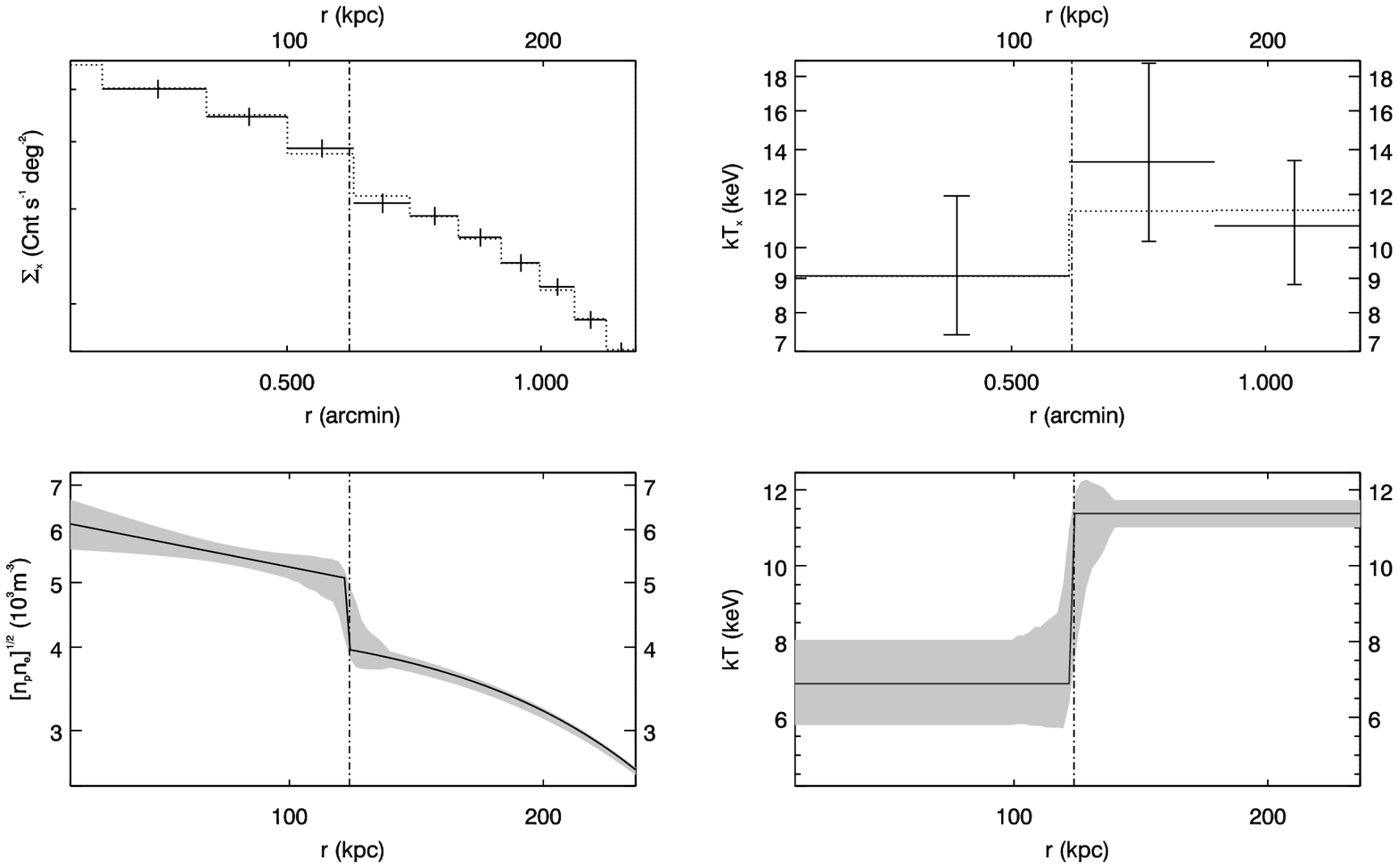}}
  \caption{Top profiles: ICM brightness (left) and projected temperature (right) profiles corresponding to the sector illustrated in the
  	bottom-left panel of \fig\ref{lxkt_maps_fig}. Bottom profiles: ICM density and 3D temperature modelled as disrupted 
	distributions with common jump position (the dot-dashed line, see also \part\ref{moving_cc} and \equ\ref{rho3d_equ} 
	and \ref{t3d_equ}). The ICM emissivity corresponding to these radial distribution has been fitted to the ICM brightness and 
	projected temperature, as shown by dotted line on the top profiles.\label{cf_fig}}
\end{figure*}

  \subsubsection{Foreground and background spectral contributions\label{foreground_background_sct}}

The spectral `background noise' model $\nb~\b(k,l,e)$ includes not only contributions related to the instrumental noise, but also those due to the extended emission of Galactic foregrounds and the Cosmic X-ray background. Following the approach described in BM08, we model the Cosmic X-ray background with a power law of index $\gamma = 1.42$ \citep[see e.g.][]{Lumb_02} and the Galactic foregrounds by the sum of several absorbed and unabsorbed thermal components. 

As already noted by \citet{Pratt_01} and \citet{Markevitch_01}, the Galactic foreground emission near \a2163 includes a strong excess in the soft band ($E < 1 \kev$), with regards to the average emission expected from `blank sky' observations. Taking into
account this excess, in addition to the local variations of $\nh$ (see Section \ref{Nh_section}) is critical to perform any spectral
analysis, in particular when measuring the ICM temperature in the cluster outskirts. As suggested by \citet{Pratt_01}, the origin of this soft excess is probably related to the location of \a2163 near the North Polar
Spur (NPS). For this reason, we decided to model the Galactic foreground emission by the sum of an unabsorbed
thermal emission accounting for the local hot bubble \citep[LHB, $\mathrm{kT}_{LHB}=0.1~\kev$, see e.g.][]{Kuntz_00} and three 
absorbed thermal components accounting for the Galactic `transabsorption emission' \citep[TAE, $\mathrm{kT}_1=0.099~\kev$, 
$\mathrm{kT}_2= 0.248~\kev$, see][]{Kuntz_00} and the NPS (see \fig\ref{background_fig}). In order to constrain this complex 
spectral model, we took advantage of the sky coverage provided by the four offset pointings available in the XMM-Newton archive (see \tab\ref{exposure_fig}) and fitted the various background  
components to the X-ray emission from an external annulus ($2700<r<4000$ kpc) where no cluster emission is expected to be
detectable\footnote{This radial range corresponds to a region located well beyond $r_{200}$, as estimated from optical
  measurements in Paper I}. Modelling the spatial variations of Galactic $\nh$ as detailed in Section
\ref{Nh_section}, and fixing the temperature of the thermal components representing the galactic TAE and LHB enabled us to
constrain the emissivity and temperature of the excess component. The temperature value obtained ($\kt_{NPS} \simeq 0.3 \kev$) is in agreement  with expectations for emission from the North Polar Spur \citep[see e.g.][]{Willingale_03}.

\section{ICM dynamics and merger events \label{icm_dynamics}}

\subsection{Gas distribution and thermal structure of A2163}

The Chandra-ACIS brightness map and iso-contours in the top left and right panels of \fig\ref{lxkt_maps_fig} show us the complex and irregular morphology of the ICM in \a2163. The map and contours reveal a main component (A2163-A) and a Northern subcluster (A2163-B), the main component itself appearing irregular with an eastern extension 
on a large scale and a triangular shape in the central region. The high resolution residue obtained from wavelet analysis of the Chandra-ACIS `soft' image --as detailed in \part\ref{lxkt_maps} and shown in the bottom-left panel of \fig\ref{lxkt_maps_fig}-- enables us to resolve further details in the central region, and to separate a brightness peak located close to the brightest cluster galaxy (BCG1, see also Paper I) from a sharp-edged, wedge-shaped feature located to its west.

The \xmm-EPIC temperature map in \fig\ref{lxkt_maps_fig} shows the very  hot mean temperature and anisotropic thermal structure of the ICM near the centre of the main cluster \a2163-A, and also the colder temperature ($\kt \simeq 6 \kev$) of the Northern subcluster, \a2163-B. The map exhibits in particular a hot region to the North of \a2163-A ($\kt \simeq 18 \kev$) and a colder clump to the SouthWest ($\kt \simeq 10 \kev$). This overall thermal structure is consistent with former analyses of Chandra data by e.g. \citet{Govoni_04}, \citet{Million_09} or \citet{Owers_09}, apart from slight morphological differences in the shape and location of the hot region in our maps. This hot region is indeed spatially coincident with a prominence of Galactic IR emissivity, likely leading to a region of enhanced absorption  (see \fig\ref{nh_map_fig}). The better sensitivity of the XMM-Newton mosaic further enables us to detect a clear drop of the ICM temperature toward the outskirts of \a2163-A (r $>$ 700 kpc). 

\subsection{A gas `bullet' crossing the main cluster atmosphere separately from galaxies\label{moving_cc}}

In the bottom-right panel of \fig\ref{lxkt_maps_fig} we show the high resolution residue obtained from the Chandra analysis overlaid on the \xmm-EPIC temperature map. Interestingly, we notice that the SouthWestern cold clump in \a2163-A is associated with the wedge-like feature revealed from the Chandra data. The cold clump thus appears as a cool core embedded in the hotter ICM of \a2163-A, that is delimited in the westward direction by a brightness edge. 

In order to investigate 3D structure of this cool core, we extracted the brightness and temperature profiles corresponding to an ICM sector incorporating the brightness edge, as illustrated in the bottom-left panel of \fig\ref{lxkt_maps_fig}. As shown on \fig\ref{cf_fig}, these two profiles have enabled us to model the ICM density $\rho(r)$, and 3D temperature $\mathrm{T}(r)$, as two disrupted functions defined in \equ(\ref{npne_cf_equ}) and (\ref{t3d_cf_equ}), with common jump location and amplitudes. The density and temperature jump amplitudes thus measured are $D_{n} = \frac{n_1}{n_2} = 1.28^{+0.07}_{-0.06}$ and  $D_\mathrm{T} = \frac{\mathrm{T}_2}{\mathrm{T}_1} = 1.65^{+0.21}_{-0.14}$, respectively, where indices 1 and 2 refers to internal and external core regions. Consistent with continuous pressure at the jump location ($D_\mathrm{P} = D_{n} D_\mathrm{T} = 1.29^{+0.20}_{-0.13}$), these jumps are likely to be associated with a cold front. This cold front has also been seen in the same data set by \citet{Owers_09}, who measured a similar pressure jump to ours from analysis of a region corresponding to the Southern part of our sector where the density jump appears as more prominent in projection. By analogy with 1E0657-56 (the so-called `bullet-cluster', \citealt{Markevitch_02}), the location of the cold front with respect to the wedge-shaped residual emission might indicate the westwards motion of a stripped cool core embedded in the hotter atmosphere of \a2163-A.

To try and constrain the history of a possible subcluster infall associated with this cool core, in the bottom-right panel of \fig \ref{lxkt_maps_fig} we overlay the galaxy iso-density contours obtained from WFI data (Paper I) on the ICM temperature map of \a2163. As discussed in Paper I, the galaxy iso-density contours map a complex, bimodal galaxy distribution in \a2163-A. This bimodal distribution corresponds to an E-W elongation of the cluster dark matter halo, as revealed from weak lensing analysis of \citet{Radovich_08}. Interestingly, we now observe that the less dense of the two galaxy subclusters in A2163-A is located near the cool core, but is  separated from it at a projected distance of about 30 arcsec. Its location  to the west of the cool core suggests a scenario in which a subcluster has crossed the \a2163-A complex from east to west; the offset wedge-shaped residual emission then suggests subcluster gas loss gas due to ram pressure of the main cluster atmosphere, leading to its present separation from the constituent galaxies.

\begin{figure}[h]
  \resizebox{.95\hsize}{!}{\includegraphics{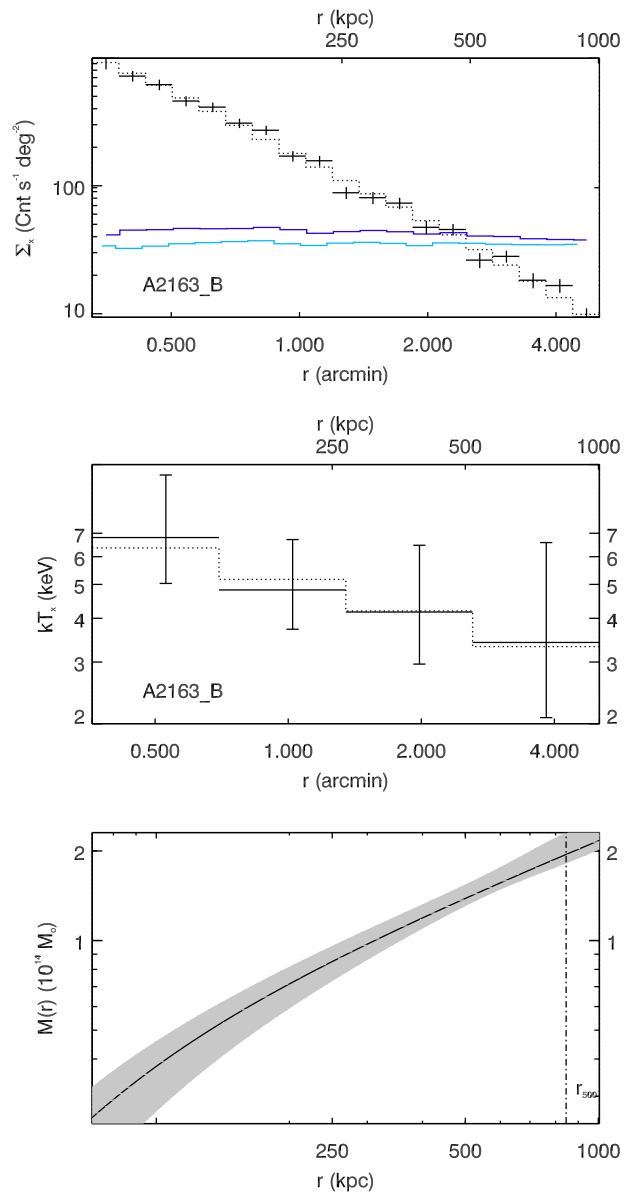}}
  \caption{\a2163-B. Top: ICM surface brightness profile. Light blue: blank-sky background emissivity. Blue: blank-sky 
        background and main cluster emissivity. Middle: projected ICM temperature profile. Bottom: Mass profile modelled from 
        the surface brightness and projected temperature profiles, assuming ICM hydrostatic equilibrium. The ICM surface 
        brightness and projected temperature have been modelled analytically (see dotted lines on the corresponding plots). \label{a2163b_prof_fig}}
 \end{figure}

\subsection{The subcluster \a2163-B}

\subsubsection{\a2163-B. Gas properties and mass}

Located at close projected distance ($\simeq$ 1.25 Mpc) from the main cluster \a2163-A, \a2163-B appears in Fig.~\ref{lxkt_maps_fig} as a small subcluster with regular morphology, hosting colder gas than \a2163-A. Due to the small angular distance separating \a2163-B from its bright companion, the X-ray emission from both \a2163-A and \a2163-B must be analysed simultaneously. We separated the two cluster emissions by modelling the emissivity from \a2163-A in each energy band from profiles $\rho(r)$ and $\T(r)$, used to model its gas density and temperature structure in Sect. \ref{mass_sct}. To do so, a 2D $\beta$-model was first fitted to the surface brightness distribution of \a2163-A, enabling us to derive ellipticity parameters and to correct the projected emissivity, $\Sigma(d,e) = 2 \int_0^{\infty} \frac{\rho[r,\T(r),e]}{\sqrt{r^2-d^2}} r dr$, for these parameters. In order to further reduce the contribution from \a2163-A, the spectra from \a2163-B have been extracted in a northern semi-sector centred on the subcluster emission peak and extending away from the main cluster emission.

This spectral component separation has enabled us to derive the residual brightness and temperature profiles of \a2163-B. As shown in \fig\ref{a2163b_prof_fig}, the gas temperatures obtained are lower than expected from visual inspection of the temperature map of Fig.~\ref{lxkt_maps_fig}, since here the emission contribution from the main cluster has been removed.  
Fitting this data set with analytical density and temperature profiles has enabled us to derive a mass profile of \a2163-B, 
assuming hydrostatic equilibrium. The cluster gas and total mass estimates derived from this analysis are $\mathrm{M}_{g,500} = 0.22^{+0.02}_{-0.03} \times 10^{14} \msol$ and $\mathrm{M}_{500} = 2.1 \pm 0.1 \times 10^{14} \msol$, respectively, where confidence intervals have been obtained from random realisations of parameters in the fitted density and temperature profile of \equ (\ref{rho3d_equ}) and (\ref{t3d_equ}). 

\subsubsection{\a2163-B. interaction with main cluster, \a2163-A}

Using galaxy velocity measurements from paper I and applying the two-body formalism, we calculated a set of bounded solutions for the dynamics of the merger system \a2163-A -- \a2163-B, with various separation distances including close solutions where both subclusters are almost observed in the sky plane, and more distant solutions where both clusters are observed as aligned along the line of sight (see Appendix for details). The ICM properties of \a2163-B do not show any strong evidence of interaction with the main cluster atmosphere. In particular, the gas brightness and temperature maps in \fig\ref{lxkt_maps_fig} do not exhibit any strong thermal or pressure enhancements in the direction of \a2163-A within an over-density radius of $\mathrm{r}_{500}$, in contrast with expectations of hydrodynamical N-body simulations of bimodal cluster mergers  \citep[see e.g.][]{Ricker_01}.
Moreover, the gas mass fraction in \a2163-B is close to the value expected for an isolated system ($f_{gas,500} \simeq 0.1$). From the lack of any close interaction evidence with \a2163-A, it is likely that \a2163-B is physically separated from its companion at a distance larger than the radius, $\mathrm{r}_{500}$, of the main cluster.

\begin{figure}[t]
    \resizebox{\hsize}{!}{\includegraphics{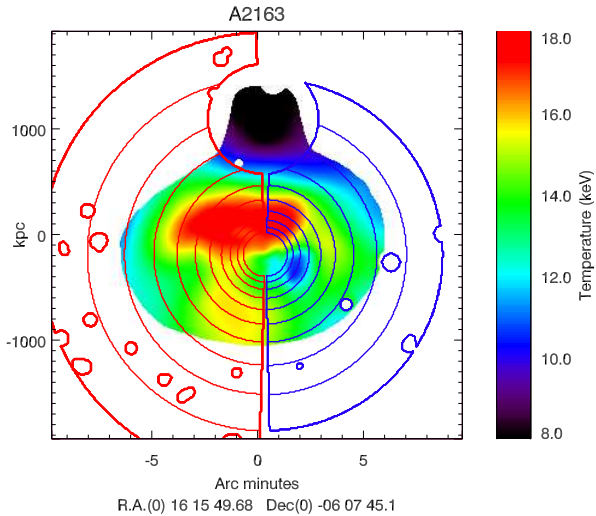}}
  \caption{\a2163 gas temperature map with two complementary profile extraction regions overlaid. The blue region is  expected to be located ahead of the moving cool core, while the red region is expected to be located behind. Both regions are centred on the X-ray emission peak\label{sector_ktmap_fig}.}
\end{figure}

\section{Merger effect on the gas pressure \label{icm_pressure_sct}}

The evidence for a gas `bullet' separated from its galaxies crossing the \a2163 atmosphere suggests that the main cluster has recently accreted a subcluster along the East-West direction (see Section \ref{moving_cc}). In order to investigate the effects of this major merger event on the ICM thermodynamics, we extracted three sets of ICM brightness, 
temperature and pressure profiles corresponding to the eastern and western cluster sectors shown on \fig\ref{sector_ktmap_fig}, and to the overall cluster \a2163-A. 

\begin{figure*}[t]
    \resizebox{.98\hsize}{!}{\includegraphics{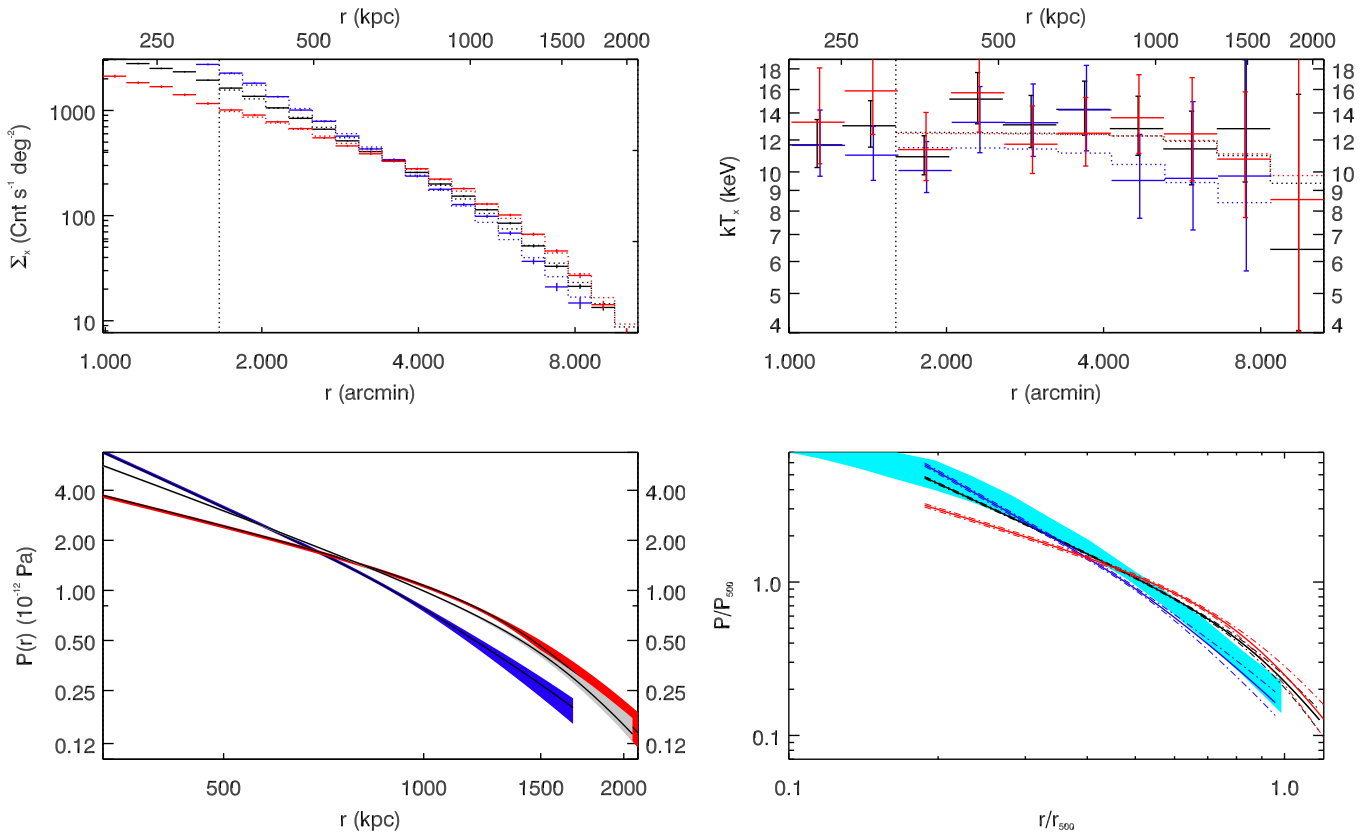}}
  \caption{\a2163-A profiles associated with the overall cluster (in black) and the eastern and western cluster sectors shown on
  \fig\ref{sector_ktmap_fig} (in red and blue, respectively). Top: Gas surface brightness (left) and projected temperature 
  profiles (right). Bottom left: ICM pressure profiles modelled to fit surface brightness and projected temperature. 
  Bottom right: ICM pressure profiles re-scaled to the characteristic pressure of a self-similar cluster with mass $\mathrm{M}_{500}$. The distribution of scaled pressure profiles measured in a representative sample of nearby galaxy clusters \citep[REXCESS, ][]{Arnaud_10} has been overlaid in light blue.\label{sector_profs_fig}}
\end{figure*}

\subsection{Brightness and temperature profiles}

Extracted from the \xmm data set and shown on \fig\ref{sector_profs_fig}, the ICM brightness and projected temperature profiles have been centred on the eastern X-ray peak revealed on the Chandra high resolution map of \fig\ref{lxkt_maps_fig}, and located close to the cluster BCG1 as shown in Paper I. This choice has enabled us to centre the profiles near the expected cluster mass centre, while radially integrating the innermost gas brightness distribution along the X-ray isophotes. The maximum radius of the profiles has been roughly set to a distance where cluster reaches the `background noise' emissivity in the soft X-ray band. Due to the strong East-West asymmetry of the cluster atmosphere we are able to extract profiles up to larger radius on the eastern side. 

We first observe that the shapes of the eastern and western brightness profiles differ strongly near the centre of \a2163-A, due to the asymmetry introduced by the crossing `bullet'. The two profiles cross each other at a radial distance of about 3.5 arcmin, beyond which the eastern brightness profile exceeds the western one. As expected, the overall cluster profile averages the radial brightness values corresponding to the eastern and western cluster sectors. Given the limited statistics available, the eastern and western temperature profiles show only marginal differences, most noticeably for the outermost radial bins and in the second radial bin intercepting the western cool core. Interestingly, we observe that the overall temperature profile seems to reflect the higher temperatures on the eastern cluster side, where the radial ICM emissivity is the highest.

\subsection{Pressure profiles}

As shown in the bottom-left panel of Fig.~\ref{sector_profs_fig}, the ICM brightness and temperature profiles have been used to model three pressure profiles corresponding to the eastern, western and overall cluster outskirts ($r \gtrsim 400$ kpc). The ICM density and 3D temperature profile have been modelled following parametric radial distribution functions --see \equ (\ref{rho3d_equ}) and (\ref{t3d_equ})--, and multiplied by each other to yield the pressure. As observed for the brightness profiles, the eastern and western pressure profiles differ clearly in the innermost region of A2163-A, then cross each other at a radial distance of about 800 kpc, above which the eastern profile has a higher pressure than  the western profile by a factor $\sim 2$. This pressure excess may be related to ICM shock heating behind the crossing `bullet', leading to an adiabatic compression of the gas. The overall cluster pressure profile seems to reflect the higher pressure values corresponding to the eastern cluster sector, so that the ongoing merger event is likely to have strongly disturbed the shape of the overall cluster pressure profile.

A recent analysis of a representative sample of nearby galaxy clusters \citep[REXCESS, ][]{Bohringer_07} has shown that the ICM pressure is expected to follow an average profile with low dispersion partially related to the thermodynamical state of clusters \citep[][hereafter A10]{Arnaud_10}. In order to be compared with average cluster pressure profile of the REXCESS sample, the ICM pressure profiles of \a2163-A have been rescaled to characteristic pressures ${\rm P}_{500}$ related to the cluster mass ${\rm M}_{500}$, in the framework of self-similar cluster evolution models (see Eqn. 5 of A10). As detailed in the following Section, the cluster mass we used for this scaling was estimated from the $Y_X$ proxy. The bottom-right panel of \fig\ref{sector_profs_fig} shows that the scaled profiles associated with the overall cluster and the western sector both follow the universal shape of the REXCESS cluster sample, while the scaled profile associated with the eastern sector appears shallower. Consistent with the discrepancy observed by A10 between relaxed and morphologically disturbed clusters in the REXCESS sample, this trend favours a scenario where the western side of \a2163-A shows the pressure structure of a self-similar cluster, while the pressure structure in the eastern cluster side has been strongly disturbed by the ongoing merger event.
\begin{table*}[ht]
\caption{\a2163 gas and total mass estimates performed from Yx mass proxy calibrated according to \citet{Arnaud_10}, \citet{Vikhlinin_09}, hydrostatic equilibrium assumption in the overall cluster ($M_\mathrm{500,HE}$), western ($M_\mathrm{500,HE,West}$) and eastern ($M_\mathrm{500,HE,East}$) cluster sectors of \fig\ref{sector_ktmap_fig}. 
Confidence intervals in the Yx mass estimates include measurements uncertainties on gas mass and temperatures, 
and systematic uncertainties on the Yx-$M_\mathrm{500}$ scaling relations. Uncertainties on hydrostatic mass estimates have been obtained from random realisations of parameters in the fitted density and temperature profile of \equ (\ref{rho3d_equ}) and (\ref{t3d_equ}).
\label{a2163_mass_tab}}

\begin{center}
\begin{tabular}{cccccc}
\hline\hline
            & $\yx,\mathrm{XMM}$ (Arnaud et al., 10) &  $\yx,\mathrm{Chandra}$  (Vikhlinin et al., 09) & HE & HE, West                &  HE,East    \\
\hline
$\mathrm{M}_\mathrm{500} (10^{14} \msol) $      &  $18.7^{+2.0}_{-2.4}$  & $20.4^{+2.9}_{-3.5}$  & $24.7^{+0.5}_{-0.9}$ & $14.5^{+0.6}_{-0.3} $ & $26.5^{+1.9}_{-1.2} $ \\
$\mathrm{M}_\mathrm{g,500} (10^{14} \msol) $   &  $2.74 \pm .01$           & $2.84\pm.01$            & $3.10^{+0.5}_{-0.9}$ & $2.54^{+0.10}_{-0.05}$ & $3.17^{+0.20}_{-0.13}$ \\
$r_\mathrm{500} (kpc) $     &  $1752^{+65}_{-72}$   & $1805^{+88}_{-99}$    & $1930^{+36}_{-70}$ & $1601^{+61}_{-29}$   & $1967^{+ 113}_{-72}$ \\
\hline
\end{tabular}
\end{center}
\end{table*}

\section{Mass measurements\label{mass_sct}}

\begin{figure}[ht]
    \resizebox{\hsize}{!}{\includegraphics{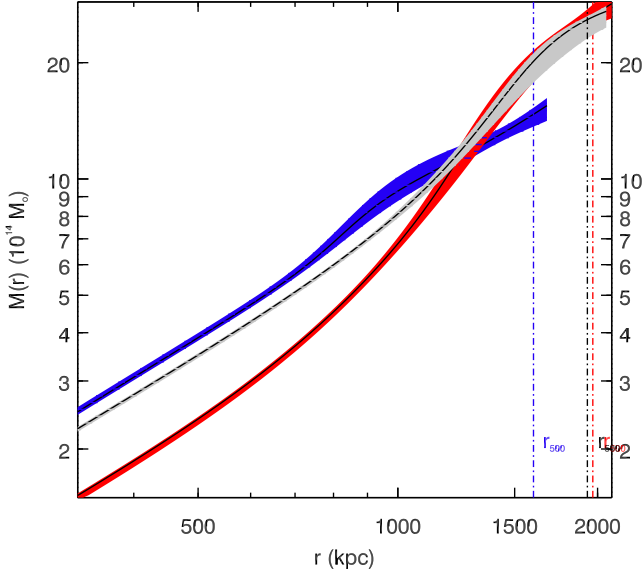}}
  \caption{\a2163-A mass profiles computed assuming ICM hydrostatic equilibrium. Profiles in the overall cluster region (in black), and the eastern and western cluster sectors illustrated in \fig\ref{sector_ktmap_fig} (in red and blue, respectively) are shown. The profiles have been derived from analytical ICM density and temperature distributions fitted to the projected gas brightness and temperature profiles of \fig\ref{sector_profs_fig}.\label{he_mass_fig}}
\end{figure}

The exceptionally high ICM temperature of \a2163 and the evidence of an ongoing major merger event in its central region makes this cluster an interesting test
case for cluster mass measurements.  We first estimated the overall mass of \a2163 
from the $\yx$ mass proxy, then investigated the validity of the hydrostatic equilibrium assumption in the overall cluster and in specific cluster sectors.

\subsection{Mass estimate from the $\yx$ proxy}

Defined as the product of gas mass $\mathrm{M}_\mathrm{g,500}$ and average temperature $\kt$, the $\yx$ parameter has been proposed by \citet{Kravtsov_06} as a robust proxy for the total cluster mass. Related to the ICM thermal energy, the parameter has shown a tight correlation with cluster mass in mock X-ray observations of clusters evolving in a cosmological context, regardless of their dynamical state. 

We estimated the $\yx$ parameter in \a2163 iteratively, yielding the total cluster mass $\mathrm{M}_\mathrm{500}$, radius $r_{500}$, gas mas $\mathrm{M}_\mathrm{g,500}$, and average temperature $\kt$. We iterate about  the $\yx-\mathrm{M}_\mathrm{500}$ scaling relation calibrated from hydrostatic mass estimates in a nearby cluster sample observed with \xmm (A10), and compare with a similar relation obtained from Chandra data \citep[][hereafter, V09]{Vikhlinin_09}. The gas masses $\mathrm{M}_\mathrm{g,500}$ were computed by integrating the ICM density profile used to model gas pressure as detailed in the previous section, while average temperatures have been estimated within radii interval of [0.15 - 0.75] $r_{500}$ and [0.15 - 1] $r_{500}$, following the $\yx-\mathrm{M}_\mathrm{500}$ relations of A10 and V09, respectively. As shown in \tab\ref{a2163_mass_tab}, the two $\yx$ mass estimates obtained are consistent, with a lower estimate from \xmm than from Chandra due to the difference in satellite calibrations. Our \xmm cluster mass estimate from iteration about the V09 scaling relation is 7 per cent lower than that found by V09 using Chandra data. This mild difference is likely to be related to the lower average temperature estimate of the present \xmm analysis, possibly due residual instrument cross-calibrations issues and small differences in Galactic absorption modelling.
 
\subsection{Mass estimates assuming ICM hydrostatic equilibrium}

Assuming hydrostatic equilibrium, the ICM density and temperature profiles used to model gas pressure have been used to derive the integrated cluster mass profile, shown as a grey shaded area in \fig\ref{he_mass_fig}. As for the pressure measurements, this profile has been centred on the eastern X-ray peak of the Chandra map and estimated in the radius range 450--2100 kpc, excluding the central region and the crossing `bullet'. As shown in \tab\ref{a2163_mass_tab}, the total cluster mass estimate $\mathrm{M}_\mathrm{500}$ obtained from this profile is inconsistent with the estimates obtained from the two \xmm and Chandra $\yx$ proxies (27 per cent and 19 per cent higher, respectively), possibly due to a departure of the ICM from hydrostatic equilibrium.

To investigate the origins of this inconsistency, we derive two additional integrated cluster mass profiles using the density and temperature profiles from the eastern and western cluster sectors shown in \fig\ref{sector_ktmap_fig}, again using analytical models and assuming ICM hydrostatic equilibrium. As shown in \fig\ref{he_mass_fig}, the eastern and western cluster mass profiles significantly differ in shape. We observe that the western profile has a higher integrated mass than the eastern profile near the cluster centre but then flattens at larger radius, leading to a lower integrated mass estimate in the cluster outskirts. 
Similar to pressure profiles in \fig\ref{sector_profs_fig}, we further observe that the overall cluster mass profile appears to reflect the mass in the eastern cluster side. \a2163 being a complex cluster system, we cannot exclude the inconsistency between eastern and western profiles to reflect a true anisotropy in the cluster mass distribution. The hypothesis of a strong mass excess beyond 1000 kpc in the eastern cluster side is however difficult to reconcile with the lack of any observational evidence for prominent galaxy or dark matter over-density in this cluster region --see paper I and \citet{Radovich_08}--. The inconsistency observed between eastern and western profiles may instead show us that the shock heating of the cluster ICM induced by the  accretion may not only have adiabatically compressed the eastern cluster atmosphere, but also transiently taken gas away from hydrostatic equilibrium in this region. For this reason, the cluster mass estimates assuming hydrostatic equilibrium in this region reported in \tab\ref{a2163_mass_tab} and hence in the overall cluster are likely to be overestimated. Assuming hydrostatic equilibrium in the western cluster side might instead provide with us an estimate of the cluster mass before the ongoing merger event, $M_{500,HE,West} = 1.5\times10^{15}~\msol$, while the subcluster accretion might have raised the total cluster mass up to values provided by the $\yx$ proxy: $M_{500,\yx} \simeq 1.9\times10^{15}~\msol$. 

\section{Discussion and conclusions \label{conclusion}}

The \a2163 system has been known to host a massive and elongated galaxy-cluster, \a2163-A, 
and a smaller associated subcluster, \a2163-B. We provided new mass measurements of the two companion subclusters, and estimate that with a mass of about $M_{500} \simeq 2.1\times10^{14}~\msol$, \a2163-B is ten times less massive than its extended companion.  As the ICM properties of \a2163-B do not show any strong evidence of interaction with \a2163-A, the two subclusters \a2163-A are likely to be physically separated at a distance larger than main cluster radius ($\mathrm{r}_{500} \simeq$ 2 Mpc). Since the two subclusters were suggested in Paper I to be connected by a galaxy bridge, they might form a bounded system observed along the line of sight, consistent with large separation solutions in the two-body dynamical analysis detailed in Appendix.

From evidence of a stripped cool core crossing the main cluster atmosphere separately from a nearby galaxy 
over-density, we infer that \a2163 might have recently accreted a subcluster along its East-West elongation. 
We suggest that this merger event has shocked the main-cluster atmosphere, and adiabatically compressed the ICM
behind the crossing cool core. This indication of shock heating and the evidence of a galaxy-gas 
separation lead us to infer that the subcluster has been accreted to a supersonic velocity. Shock heating may further have
transiently pulled cluster gas away from hydrostatic equilibrium, leading to large uncertainties on the cluster mass estimate from X-ray analysis. 
Assuming the $\yx$ proxy to indicate the mass of the overall system \a2163 and the western cluster sector to represent the ICM in hydrostatic equilibrium before subcluster accretion, we may infer a merger scenario between two subunits of mass-ratio 1:4, leading to the present system with mass $M_{500} \simeq 1.9\times 10^{15}~\msol$.

A fast subcluster accretion with evidence of galaxy-gas separation was first seen in 1E 0657-56, the so-called `bullet cluster' by \citet{Markevitch_02}.  It is interesting to note that both 1E 0657-56 and \a2163 are exceptionally massive, favouring ram pressure stripping of the gas content of an incoming subcluster due both to the high density of these accreting cluster atmospheres, and by exceptionally high collision velocities assuming free fall encounters. It is thus likely that the comparable mass of A2163 and 1E 0657-56 has 
favoured similar merger scenarios. Moreover, both 1E 0657-56 and \a2163 exhibit similar global properties such as exceptionally high luminosity and gas temperature (about 15 and 12 keV for 1E
0657-56 and \a2163, respectively), and powerful emission from an
extended radio halo. These common properties are likely point to the recent dissipation of a large amount of kinetic energy through shock heating and turbulence.

While presenting striking similarities related to 
a similar `bullet' accretion scheme, \a2163 and 1E 0657-56 show different morphologies that may point to some small differences between time
scales and the initial conditions of the accretion process. The crossing
`bullet' is observed at a larger projected cluster radius in
the case of 1E 0657-56 (500 kpc vs. 290 kpc as for 1E 0657-56 and
\a2163, respectively), and is thus better separated from the dense
emission of the main cluster. Assuming both `bullet' motions to be roughly
orthogonal to the line of sight, the projected distances would
represent physical distances to the cluster core, and the `bullet'
would be observed at an earlier point in its crossing of the main cluster in the case of \a2163. Unlike for the 1E 0657-56 `bullet', however, the \a2163 `bullet' does not seem to be preceded by a clear shock front. This 
non detection might be the result of a less favourable merger geometry, 
with a larger angle separating the \a2163 collision axis from the sky plane.
As discussed in Paper I, the galaxy velocity distribution in the main cluster indeed shows a strong velocity gradient of $\sim 1250$ km/s along a NE/SW axis, which is much higher than the velocity offset of $\sim 600$ km/s separating the two galaxy components of 1E 0657-56 \citep{Barrena_02}. 

As extensively studied in the textbook case of the `bullet cluster', observing a high-velocity subcluster accretion within a massive cluster may provide interesting constraints on
ICM physics, galaxy evolution, and dark matter properties. Better constraints on the dynamics of the exceptional merger event in A2163 would thus be obtained from deeper X-ray or SZ observations, allowing us e.g. to constrain the `bullet' velocities from precise pressure 
measurements at the cold front stagnation point, or to look for a shock front ahead of the `bullet' direction of motion. Moreover, 
the present observations might be compared to a high resolution weak lensing analyses, enabling us to verify if the bimodality of the galaxy distribution seen in Paper I reflects the underlying dark matter morphology, 
and whether the galaxy-gas separation presently observed also corresponds to a segregation between the baryonic and dark matter content of the accreted subcluster.

\begin{acknowledgements}      

We thank the referee for constructive remarks that have helped us to improve our manuscript. H.B. and P.M. 
acknowledge financial support from contracts ASI-INAF I/088/06/0, NASA GO8-9126A and GO9-0145A. C.F. 
acknowledges financial support by the Agence Nationale de la Recherche through grant ANR-09-JCJC-0001-01.
This work is based on observations obtained with XMM-Newton, an ESA science mission funded by ESA 
Member States and the USA (NASA).

\end{acknowledgements}

\bibliography{a2163_aa3}

\section*{Appendix: Two-body analysis of A2163A/A2163B}

\begin{figure*}
\centering
\resizebox{\textwidth}{!}{\includegraphics{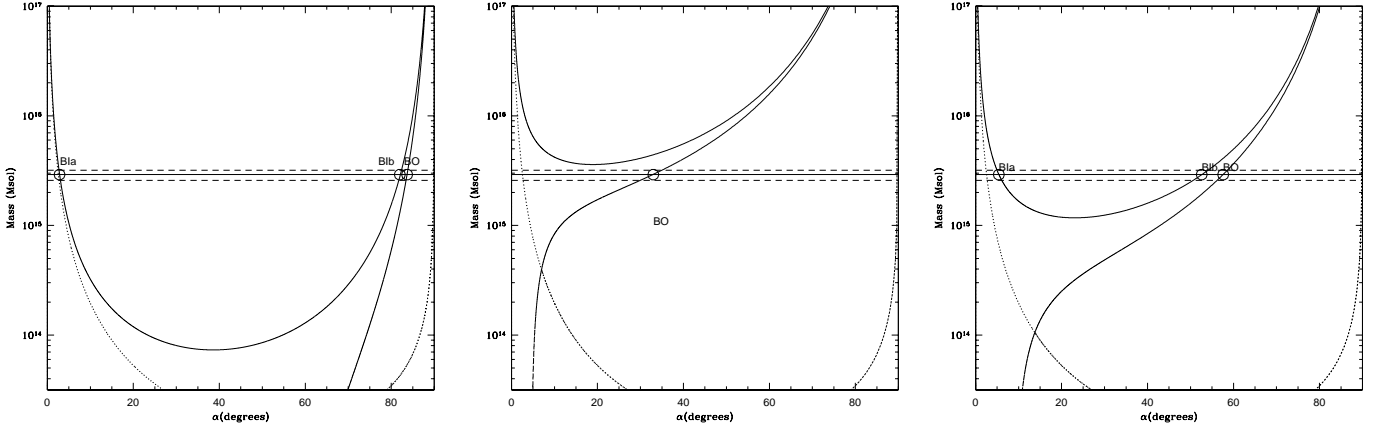}}
\caption[]{The sum of the virial masses of A2163-A and A2163-B as a function of the projection angle. The horizontal lines show the cluster mass estimate (full line) and errors (dashed lines) obtained in the present paper. The projected distance of 1.25 Mpc and radial velocity difference of 200 km s$^{-1}$ are derived from Maurogordato et al. (2008). The two systems were at zero separation 10 Gyr ago (left panel), 0.5 Gyr ago (middle panel), and 1.0 Gyr ago (right panel). The dotted curve corresponds to the Newtonian criterion for gravitational binding. Bound solutions are above this plot.}
\label{fig::mass_alpha}
\end{figure*}

\begin{table}[ht]
\begin{center}
\small
\begin{tabular}{cccccc}
\hline
\hline
t        &solution  &$\alpha$(deg) & R(Mpc)  & Rm(Mpc)  &V(km/s)\\
\hline
10 Gyr    &BIa       &2.83        &1.25     &6.5       &4050\\
          &BIb       &82         &9.1    &9.3      &200\\
          &BO        &84         &11.4     &11.6      &200\\    
\hline
0.5Gyr    &BO        &33        &1.49    &1.5      &370\\ 
\hline
1.0Gyr    &BIa       &5.4       &1.25    &1.5      &2100\\
          &BIb       &52.6       &2.06    &2.07      &250\\
          &BO        &57.6       &2.33     &2.34      &240\\
\hline
\hline
\end{tabular}
\caption{Solutions of the two body model for the A2163-A--A2163-B system, for the three cases considered: before first approach (10 Gyr), 0.5 Gyr and 1.0 Gyr after first approach. For each solution, the angle $\alpha$ between the line connecting the two components and the plane of the sky $\alpha$, the spatial separation of the subclusters R, their separation at maximum expansion Rm and their relative velocity V are determined. }
\label{Table_2body}
\end{center}
\end{table}

We applied the two-body formalism \citep{Beers_82} to the system composed of A2163-A (main) and A2163-B (Northern subcluster). In this method, the two components are supposed to be coincident at t=0 and are moving apart or coming together for the first time in their history. In this framework, one can write the parametric solutions to the equations of motion, relating the elapsed time $t$, the separation $R$ and the  velocity $V$ of the two components to the total mass of the system $M$, the development angle $\chi$, and the maximum expansion radius in the bound case (the asymptotic expansion velocity $V_\infty$, in the unbound case). In the case of bound radial orbits:

\begin{equation}
R = \frac{R_m}{2}~(1-\cos~\chi)
\end{equation}
\begin{equation}
t = \left(\frac{R_m^3}{8GM}\right)^{1/2} (\chi-\sin~\chi)
\end{equation}
\begin{equation}
V = \left(\frac{2GM}{R_m}\right)^{1/2}\frac{\sin~\chi}{(1-\cos~\chi)}
\end{equation}

\noindent In the unbound case:

\begin{equation}
R = \frac{GM}{V_{\infty}^2}~(\cosh~\chi-1)
\end{equation}
\begin{equation}
t = \frac{GM}{V_{\infty}^3}~(\sinh~\chi-\chi)
\end{equation}
\begin{equation}
V = V_{\infty}\frac{\sinh~\chi}{(\cosh~\chi-1)}
\end{equation}

\noindent The velocity and spatial separation are directly related to the observed  line of sight velocity difference $V_r$ and projected separation $R_p$ by:

\begin{equation}
R = \frac{R_p}{\cos \alpha}; V = \frac{V_r}{\sin \alpha}
\end{equation}
where $\alpha$ is the angle between the axis joining the two units and the plane of the sky. When the two units are bound together, the system fulfils the Newton criterion:

\begin{equation}
{V_r}^2 R_p \leq ~2GM ~\sin^2~{\alpha}~ \cos~\alpha
\end{equation}

\noindent Specifying  the angular separation, the velocity offset of the two systems, and the time, one can close the system and derive the evolution of the angle $\alpha$ as a function of these observed values and the development angle $\chi$ \citep{Owers_09}. This leads in the bound case to:

\begin{equation}
\tan \alpha = \frac{V_r t}{R_p} \frac {(1-\cos~\chi)^2} {\sin \chi (\chi-\sin \chi)}
\end{equation}
and similarly in the unbound case to: 
\begin{equation}
\tan \alpha = \frac{V_r t}{R_p} \frac {(\cosh~\chi - 1)^2} {\sinh \chi (\sinh \chi - \chi)}
\end{equation}

\noindent These equations can be used in both cases to derive the 
total virial mass of the system as a function of the projection angle $\alpha$ with respect to the plane of the sky. Comparing to the current estimates of the mass will allow us to assess which solutions are acceptable \citep{Barrena_02}. 

We have taken $R_p=1.25$ Mpc and $V_r=200 $km/s (from Paper I), and set $t$ to 10 Gyr (the age of the universe at $z=0.2$). We used the mass limits by summing mass estimates of the two components \a2163-A and \a2163-B as measured in the present paper and assuming $\mathrm{M}_{200} = 1.4 \mathrm{M}_{500}$. Fig.~{\ref{fig::mass_alpha} shows that in the pre-merger hypothesis, three bound solutions do exist: two bound incoming and one bound outgoing  solutions (see Table\ref{Table_2body}). The unbound solutions are excluded. 

We have also investigated the possibility that the first encounter between both systems has already occurred, making several trials from $t=0.1$ Gyr to 1Gyr. Recent encounters ($t<0.2$ Gyr) are not compatible with the system being bound. Taking $t=0.5$ Gyr after the presumed encounter, the only possible solution is the bound outgoing  solution. For higher values of $t$, one finds again three bound solutions (two incoming and one outgoing). 

\end{document}